\documentclass[a4paper,12pt]{article}
\usepackage[noadjust]{cite}
\makeatletter
\def\ps@headings{%
\def\@oddhead{\mbox{}\scriptsize\rightmark \hfil \thepage}%
\def\@evenhead{\scriptsize\thepage \hfil \leftmark\mbox{}}%
\def\@oddfoot{}%
\def\@evenfoot{}}
\makeatother
\usepackage{graphics}
\usepackage{graphicx}
\graphicspath{{../pdf/}{../png/}}
\usepackage{multirow}
\usepackage{amsmath}
\usepackage{booktabs}
\usepackage{xcolor}

\title{\bf Segmentation of Encrypted Data} 

\author{Eric J\"arpe$^\ast$ Quentin Gouchet$^{\ast\ast}$\\
\\
\normalsize{$^\ast$School of Information Science, Halmstad University,}\\
\normalsize{P.O.\ Box 823, 301 18 Halmstad, Sweden}\\
\normalsize{$^{\ast\ast}$Atsec Information Security, Austin, TX 78759, USA}\\
\\
\normalsize{$^\ast$To whom correspondence should be addressed; E-mail:  eric.jarpe@hh.se.}
}

\date{2014}

\begin{document} 

\maketitle 

\begin{abstract}
The retrieval of data from computer hard drives that have been seized from police busts against suspected criminals are sometimes not straight forward. Typically the incriminating data, which may be important evidence in subsequent trials, is encrypted and quick deleted. The cryptanalysis of what can be recovered from such hard drives is then subject to time-consuming brute-forcing and password guessing. To this end methods for accurate classification of what is encrypted data and what is not is of the essence. Here a procedure for discriminating encrypted data from non-encrypted is derived. Several methods are suggested and their accuracy is evaluated in different ways.

Two methods to detect where encrypted data is located in a hard disk drive are detailed using passive change-point detection. The measures of performance of such methods are discussed and a new property for evaluation is suggested. The methods are then evaluated and discussed according to the new performance measure as well as the standard measures.
\end{abstract}


\section*{Introduction}
\paragraph*{Background}
Being able to detect encrypted files may be primordial in several cases such as in police investigations where evidence of criminal activity is residing as data on a computer hard disk drive (HDD). This is a major issue in heavy criminality and organized crime \cite{polisrapport1}, \cite{MSBrapport1}. When the police seize an HDD containing data belonging to a suspected criminal, that data can be material of evidence in a subsequent trial. But criminals often try to make sure that the police will be unable to use that data. Possible actions to obstruct data access are then to encrypt and/or to {\em quick delete} that data. In case of quick delete of data the pointers to the files are destroyed but the contents of the files is still left. The other action is to encrypt files. Using state of the art tools for encryption of files (such as Bitlocker and the thereby provided ciphers) there is no general procedure of breaking the encryption other than brute force attack by guessing the cipher algorithm and systematically guessing the encryption key. If some of the files are encrypted and others not, it is then a delicate matter to distinguish between the two if the data has been quick deleted. Nevertheless, the importance of discriminating between encrypted data (i.e.\ ciphertext) and non-encrypted data (i.e.\ plaintext) is also stressed by the fact that brute force attacking all clusters of data on the hard drive would be an impossible task while being able to separate the encrypted data from the non-encrypted would make a substantial improvement of the chances for successful cryptanalysis. The police authorities usually have software to brute force those encrypted data but this procedure may be very time consuming if the amount of data is so large that different parts of it have been encrypted with different keys. In juridical cases, time is typically of the matter since the chances of success in prosecution and proceedings against a criminal is depending on deadlines (such as time of arrest, time of trial etc) any time savings in the procedure of extracting evidence from the hard drives is essential. Thus time has to be spent on the appropriate tasks: code-breaking only on the encrypted data rather than try to decipher data which are not encrypted. Given a single file,  the task of determining whether it is encrypted or not is usually easy; but given a whole hard disk drive (HDD) without knowing where the encrypted files are, this is trickier.

There are several software solutions for certifying whether a file is encrypted or not, mostly checking the header of the file and looking for some known header like the EFS (Encryption Files System on Windows), BestCrypt, or other software headers. Such alternatives cannot be used in the case when the user has performed a (quick) delete of the HDD because then the pointers to all files are lost. Nevertheless, the files actually remain on the HDD: upon deleting a file on an HDD, the pointer to the beginning and end of the physical space on the HDD containing the information of the file are removed. But the physical space on the HDD containing the information of the file is not overwritten because that operation would be slow, i.e.\ as slow as writing the same file again. The operating systems designers decide to leave the file in the HDD intact but indicates that this location is free to host some other data. If nothing has been overwritten, this means that the file is actually still stored in the HDD for some time which allows for recovery software to restore this file.

Recovery software might help to simply recover the but might also overwrite the data contained in the HDD which might result in loss of evidence in case of an investigation. In this case, the police would have to locate the encrypted files without using any recovery or "header-detector" software. In spite of being a pertinent problem in urgent need of a solution, surprisingly little has been done to investigate the properties of methods for the detection of encrypted data and comparing these to whatever is done. An attempt in this direction is \cite{DetectKrypt} but this is just a small first step.

Here, methods to locate quick deleted encrypted data are presented and detailed. First, a description of how encrypted data is different from other data is presented. This is followed by the introduction to statistical change-point detection methods for discrimination between encrypted and non-en\-cryp\-ted data. Finally, the results of these procedures are presented along with some experimental values to evaluate the methods.

However, such a method will only work on mechanical HDDs and not with flash memory devices: in flash memory (like USB memory sticks or Solid State Drive (SSD)), as soon as a file is removed it is actually erased from the memory because data cannot be overwritten. Therefore as soon as data is deleted, the operating system will choose to delete the pointers. But erasing data in a flash memory also takes longer time since all the file contents have to be removed which takes as long as copying new files to the device.

\paragraph*{Description}
If encrypted data is not uniformly distributed, the cryptosystem used to cipher those data has a bias and is in this sense vulnerable to cryptanalysis attacks. For this reason characters of the ciphertext produced by any modern high-quality cryptosystem is uniformly distributed 
\cite{DetectKrypt}, \cite{Stegano} i.e.\ the values of the bytes of the ciphertext are uniformly distributed on some character interval. The unencrypted files do not possess this feature although some types of files are actually close to having their characters being uniformly distributed. The files coming closest to uniformly distributed contents without being encrypted are compressed/zipped files: those files are indeed very close to cipher files in terms of the distribution of their character's byte numbers. Albeit small there is a difference in distribution making it possible to tell compressed files and encrypted files apart. This paper is all about how to detect such small but systematic differences and consequently how to quickly and accurately segment encrypted HDD data for efficient cryptanalysis.

\section*{Methods}
\paragraph*{Distribution of encrypted data}
% BEGIN
% OBS! OBS! OBS! OBS! OBS! OBS! OBS! OBS! OBS! OBS! OBS! OBS! OBS! 
% GET THIS RIGHT!!! GET THIS RIGHT!!! GET THIS RIGHT!!!
%
The working hypothesis is that data (i.e.\ characters) constituting encrypted files are uniformly distributed, while data of non-encrypted files are not (i.e.\ differently distributed depending on which type of non-encrypted files). The goal now is to be able to tell apart an encrypted file from a non-encrypted one.

Let us assume the data constitutes of characters divided into clusters, $c_1,c_2,c_3,\ldots$ with $N$ characters in each cluster. The characters used a range over some alphabet of a set of possible forms. Merging these forms into $K$ character classes, the counts $O_{kt}$ of occurrences of class $k$ characters in cluster $t$ are observed. One method of measuring distribution agreement is by means of a $\chi^2$ test statistic, $Q_t=\sum_{k=1}^KE_{kt}^{-1}(E_{kt}-O_{kt})^2$ where $E_{kt}$ are the expected counts of occurrences of characters in class $k$, cluster $t$. Under the hypothesis of uniformly distributed characters, the expected counts of occurrences within each class is $E_{kt}=\frac{N}{K}$. Also, by definition, $\sum_{k=1}^KO_{kt}=N$. This reduces the statistic to
\[ Q_t=-N+\sum_{k=1}^K{\textstyle\frac KN}O_{kt}^2. \]
The values of this statistic are henceforth referred to as $Q$ scores. Large $Q$ scores indicate deviance from the corresponding expected frequencies $E_{kt}$. The smallest possible $Q$ score being $0$ would be attained if all $E_{kt}=O_{kt}$. The expected value, $E_{kt}$, in each class should not be smaller than 5 for the statistic to be relevant (5 is a commonly used value; other values like 2 or 10 are sometimes used depending on how stable the test statistic should be to small deviances in tail probabilities). Therefore one should use at least $5$ kB of data in each cluster to enable this test to be relevant. But the larger the number of bytes that are in each cluster, the worse the precision to detect encrypted file values: indeed if too many bytes are detected as being unencrypted (but were actually encrypted), a large amount of encrypted data will then not be detected in the procedure.

Here the alphabet used was the numbers $0,1,\ldots,255$ representing the possible values of bytes representing the characters in the data. These numbers were divided into $K=8$ classes (class 1: values of bytes in $[0,31]$ to class 8: values of bytes in $[224,255]$) and the clusters of size $N=64$ bytes making the expected values in each class $E_{kt}=8$, $k=1,2,\ldots,8$.
%Each cluster could be separately treated to decide whether it is encrypted or not but this would not be relevant since compressed data and encrypted data's distributions are so similar: there would be way to many errors in the detection, in either ways (detecting non-encrypted data as being encrypted or vice versa).
\begin{figure}
\centering
\includegraphics[width=\linewidth]{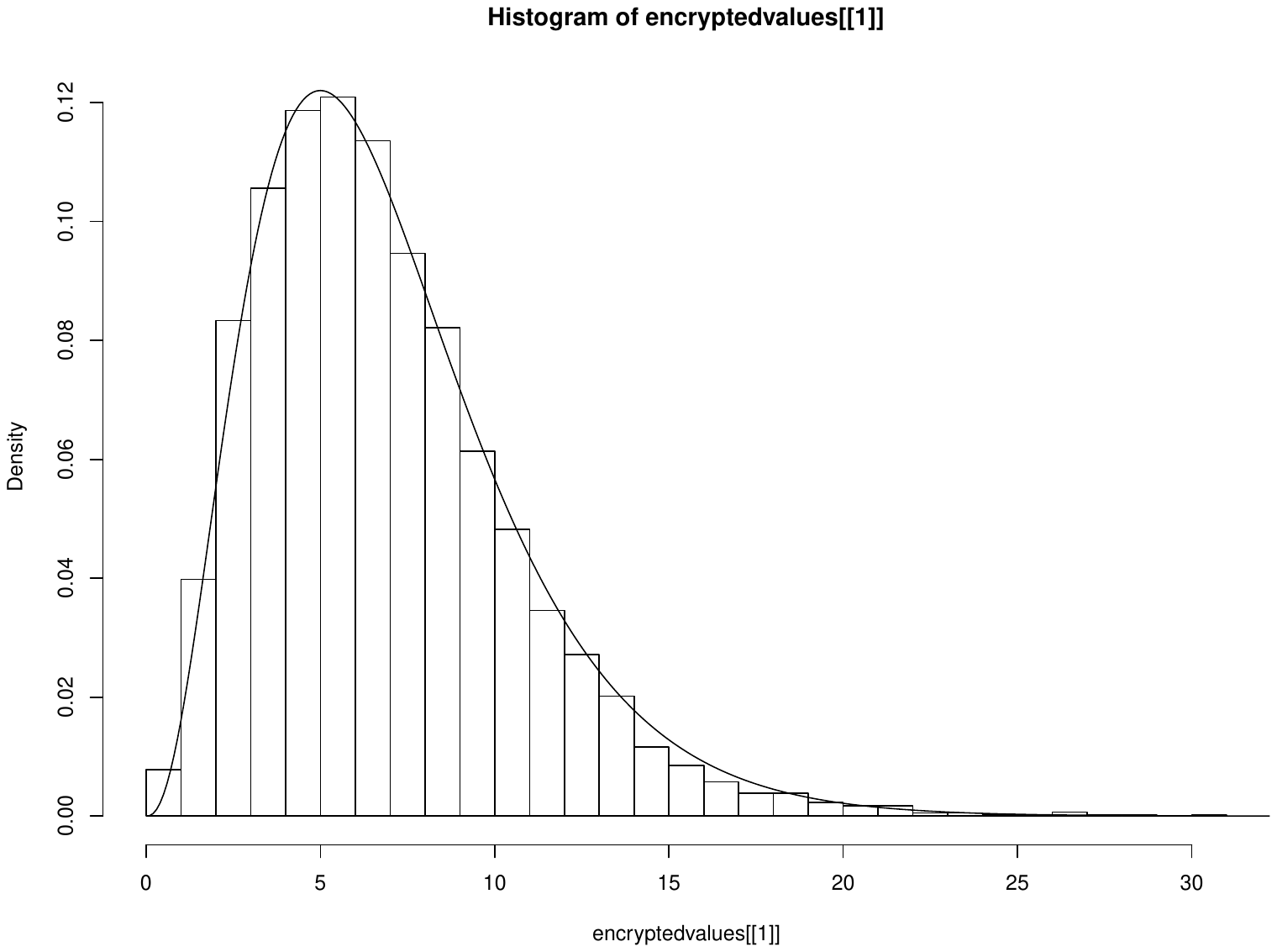}
\caption{Distribution of the $Q$ scores of encrypted files (obtained by using more than 5000 files) with the distribution function of  $Q\in\chi^2(7)$}
\label{encryptedvalues}
\end{figure}
Assuming that encrypted data is uniformly distributed, the $Q$ scores based on counts of characters in encrypted data are $\chi^2$ distributed, see Figure \ref{encryptedvalues} and since 8 classes were chosen, the number of degrees of freedom is $8-1=7$.

\paragraph*{Distribution of non-encrypted data}
For non-encrypted data, the distribution is more complicated. Basically, each type of files has its own distribution. Consequently, the standardized squared deviances from expected counts under an assumption about uniform distribution are larger and so are the $Q$ scores of the $\chi^2$ statistic. However, two problems emerge. Firstly, the size of these increased deviances depends on the type of data -- i.e.\ whether the data is a text file, an image, a compiled program, a compressed file or some other kind of file -- and how should this information be properly taken into account? Secondly, what is the distribution of the $Q$ score in the case when the data are not encrypted?

\begin{figure}
\centering
\includegraphics[width=\linewidth]{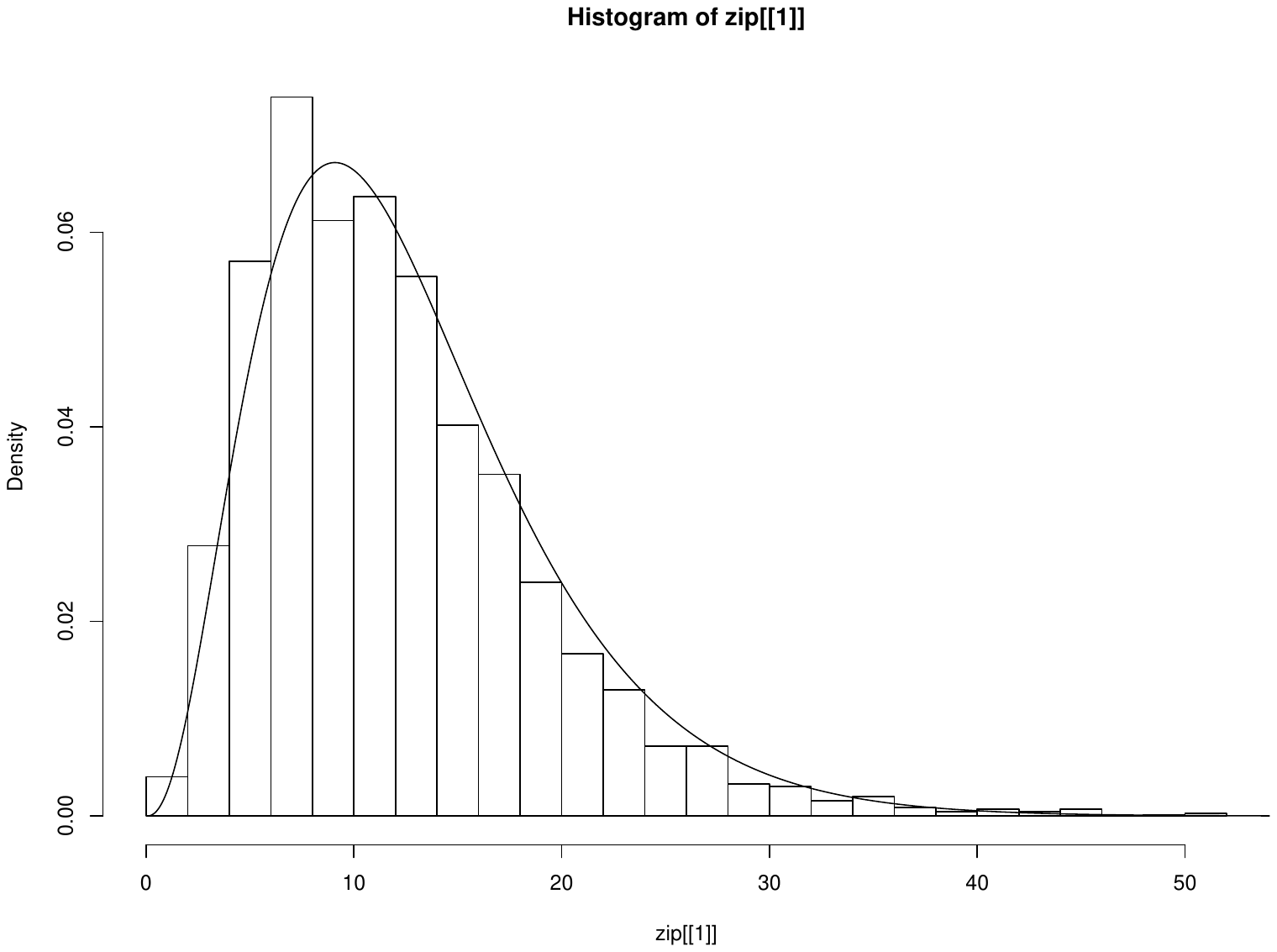}
\caption{Distribution of the $Q$ scores of compressed files (obtained by using more than 5000 files) with the distribution function of $Q=1.7374\cdot X$ where $X\in\chi^2(7)$}
\label{nonencryptedvalues}
\end{figure}

In order to develop a method for distinguishing between encrypted and non-encrypted data, it is sufficient to focus on the non-encrypted which is most similar to the encrypted and this turns out to be compressed data. Other types of files such as images, compiled programs etc.\ commonly render higher $Q$ scores and are therefore indirectly distinguished from encrypted data by a method which is calibrated for discriminating between encrypted and compressed data. About the second question, this is not readily answered. Rather we just suggest to model the $Q$ score as being scaled $\chi^2$ distributed, i.e.\ the $Q$ score is assumed to have the distribution of the random variable $\alpha X$ where $\alpha>1$ and $X\in\chi^2$. The validity of this approach is sustained by an empirical evaluation based on more than 5000 compressed files. The resulting empirical distribution of their $Q$ scores and the distribution of $\alpha X$ where $X$ is $\chi^2$ distributed and the value of $\alpha=1.7374$ was estimated by the least square method was plotted in Figure \ref{nonencryptedvalues}.
% $\alpha = 1.737388571.

\paragraph*{Change-point detection}
The object of detecting encrypted data is to quickly and accurately detect a shift in distribution from on-line observation of a random process (see e.g.\ \cite{PropShewhart,StatSurveillance,OptimalSurveillance,SurveillanceEnv}). Change-point detection can be done actively (stop collecting the data as soon as a shift is detected) or passively (continue collecting the data even if a shift is detected in order to detect more shifts). Here passive on-line change-point detection was used to detect if the data from an HDD shifts from non-encrypted to encrypted and vice versa. The change-point detection method is a stopping rule
\[ \tau=\inf\{t>0:a_t>C\} \]
where $a_t$ is called {\em alarm function} and $C$ {\em threshold}. The design of the alarm function defines different change-point detection methods while the values of the threshold reflects the degree of sensitivity of the stopping rule. The alarm function may be based on the likelihood ratio
\[ L(s,t)=\frac{f_{\mbox{\scriptsize\boldmath$Q$}_t}(\mbox{\boldmath$q$}_t\,|\;\theta=s\le t)}{f_{\mbox{\scriptsize\boldmath$Q$}_t}(\mbox{\boldmath $q$}_t\,|\; \theta>t)} \]
where $f_{\mbox{\scriptsize\boldmath $Q$}_t}(\mbox{\boldmath $q$}_t|A)$ is the conditional joint density function of the random variables $(Q_1,\ldots,Q_t)=\mbox{\boldmath $Q$}_t$ given $A$ and where $\mbox{\boldmath $q$}_t$ is the vector of the observed values of $\mbox{\boldmath $Q$}_t$. Assuming indpendence of the variables $Q_1,\ldots,Q_t$ the likelihood ratio simplifies to
\[ L(s,t)=\prod_{u=s}^t\frac{f_1(q_u)}{f_0(q_u)} \]
where $f_0(q_u)$ is the marginal conditional density function of $Q_u$ given that the shift has not occurred by time $u$ and $f_1(q_u)$ is the marginal conditional density function of $Q_u$ given that the shift occurred in or before time $u$.

%From here, $\alpha$ will denote the coefficient $1.737388571$ and $k$ the number of degrees of freedom ($k = 8-1 = 7$ in this case), $\tau$ a stopping time (when the alarm function detects a shift) and $\theta$ a real change-point time.

The conditional density function of the $Q$ score at time $t$ given that the data is encrypted (i.e.\ uniformly distributed) is
\[ f_{\mbox{\scriptsize E}}(q_t)=\frac{q_t^{k/2-1}e^{-q_t/2}}{2^{k/2}\Gamma(k/2)} \]
where $k$ is the number of degrees of freedom, i.e.\ the number of classes (which in this study is 8 as explained above).

For the non-encrypted files, the conditional $Q$ score is modelled by $\alpha X$ where $X\in\chi^2(k)$ and $\alpha>1$, supposedly reflecting the inflated deviances from the uniform distribution had the data been encrypted. Thus
\begin{equation}
    \begin{split}
      {\textstyle f_{\mbox{\scriptsize NE}}(q_t)} &= {\textstyle\frac{\partial}{\partial q_t}}P(\alpha X<q_t \;|\; X\in\chi^2(k))\\
&= \frac{(\frac{q_t}{\alpha})^{k/2-1}e^{-q_t/2\alpha}}{\alpha 2^{k/2}\Gamma(k/2)}
\end{split}
\end{equation}
is the density function of non-encrypted data $Q$ score.
This means that two cases of shift in distribution are possible:
\begin{itemize}
\item Shift from non-encrypted to encrypted data in which case
\begin{equation}
    \begin{split}
      {\textstyle L(s,t)} &= {\textstyle \prod\limits_{u=s}^t\frac{f_{\mbox{\tiny E}}(q_u)}{f_{\mbox{\tiny NE}}(q_u)}}\\
                         &= {\textstyle \alpha^{k(t-s+1)/2}\exp\Big(-\frac{\alpha-1}{2\alpha}\sum\limits_{u=s}^tq_u\Big)}.
    \end{split}
\end{equation}
\item Shift from encrypted to non-encrypted data in which case
\begin{equation}
    \begin{split}
      {\textstyle L(s,t)} &= {\textstyle \prod\limits_{u=s}^t\frac{f_{\mbox{\tiny NE}}(q_u)}{f_{\mbox{\tiny E}}(q_u)}}\\
      &= {\textstyle \alpha^{-k(t-s+1)/2}\exp\Big(\frac{\alpha-1}{2\alpha}\sum\limits_{u=s}^tq_u\Big)}.
    \end{split}
\end{equation}
\end{itemize}
To detect whether the shift in distribution has occurred or not according to the stopping rule $\tau$ mentioned above, an alarm function should be specified. Two of the most common choices here are:
\begin{itemize}
\item CUSUM \cite{page}: $a_t=\max_{1\le s\le t}L(s,t)$,
\item Shiryaev \cite{shiryaev}: $a_t=\sum_{s=1}^tL(s,t)$.
\end{itemize}
Other possible choices are e.g.\ the Shewhart method, the Exponentially Weighted Moving Avergage (EWMA), the full Likelihood Ratio method (LR) and others, see e.g.\ \cite{StatSurveillance} for a more extensive presentation of different methods.

For the CUSUM alarm function, as $\arg\max_{1\le s\le t}L(s,t) =\arg\max_{1\le s\le t}\ln L(s,t)$ the alarm function is simplified without any loss of generality by using the log likelihood values instead.
For both cases, the alarm functions can be expressed recursively which facilitates running the algorithm in practice when data streams are big.

The alarm function for shift from non-encrypted to encrypted data for the
\begin{itemize}
\item CUSUM method is
\[ a_t=\left\{\begin{array}{lll}
                0 & t=0\\
                \left(a_{t-1}+\frac{k\ln \alpha}{2}\right)^+\!+\!\frac{1-\alpha}{2\alpha}q_t & t=1,2,3,\ldots
              \end{array}
       \right. \]
\item Shiryaev method is
\[ a_t=\left\{\begin{array}{lll}
                0 & t=0\\
                \alpha^{k/2}e^{\frac{1-\alpha}{2\alpha}q_t}(1 + a_{t-1} ) & t=1,2,3,\ldots
              \end{array}
       \right. \]
\end{itemize}
The alarm function for shift from encrypted to non-encrypted data for the
\begin{itemize}
\item CUSUM method is
\[ a_t=\left\{\begin{array}{lll}
                0 & t=0\\
                \left(a_{t-1}-\frac{k\ln(\alpha)}{2}\right)^+\!+\!\frac{\alpha-1}{2\alpha}q_t & t=1,2,3,\ldots
              \end{array}
       \right.\]
\item Shiryaev method is
\[ a_t=\left\{\begin{array}{lll}
                0 & t=0\\
                \alpha^{-k/2}e^{\frac{\alpha-1}{2\alpha}q_t}(1 + a_{t-1} ) & t=1,2,3,\ldots
              \end{array}
       \right.\]
  \end{itemize}

\paragraph*{Evaluation}
To quantify the quality of different methods, the performance is compared regarding relevant properties such as the time until false alarm, delay of motivated alarm, the credibility of an alarm and so on. The threshold is commonly set with respect to the Average Run Length $\mbox{ARL}^0$ which is defined as the expected time until an alarm when no parameter shift actually occurred (which means that this is actually a false alarm). It is crucial to have the right threshold values for the methods to perform as specified. Setting the threshold such that $\mbox{ARL}^0$ is $100, 500, 2\,500$ and $10\,000$ respectively (the most common values here are $\mbox{ARL}^0=100$ and $500$ but the higher values are also considered since the value of $\mbox{ARL}^0$ defines the number of clusters/time points that are treated before a false alarm and the shift could occur very far into the HDD) properties of the methods regarding delay and credibility of a motivated alarm can be compared. Of course, if the threshold is low, this will lead to more false alarms (detection of a change when there is none) but specification with a too high threshold will lead in a drop of sensitivity of the method to detect a shift (higher delay between a shift and its detection) and consequently an increased probability of missing a real shift in distribution.

The expected delay, $\mbox{ED}(\nu)=E_\nu(\tau-\theta \;|\;\theta<\tau)$ (expectation of the delay of a motivated alarm; see Table~\ref{tab:ED}) or Conditional Expected Delay $\mbox{CED}(t)=E(\tau-\theta\;|\;\tau>\theta=t)$ (expectation of the delay when the change point is fixed equal to $t$), are important measures of performance for many applications.

\begin{table*}[t]
  \centering
\begin{tabular}{|c|c|c|c|c|c|c|c|c|c|}
 \hline
 \multirow{3}{*}{$\nu$} & \multicolumn{8}{c|}{Methods} \\
 \cline{2-9}
  & \multicolumn{4}{c|}{CUSUM} & \multicolumn{4}{c|}{Shiryaev} \\
 \cline{2-9}
  & 100 & 500 & 2\,500 & 10\,000 & 100 & 500 & 2\,500 & 10\,000 \\
 \hline
 0.2  & 4.7844 & 7.1087 & 9.5520 & 11.6905 & 4.9672 & 7.3116 & 9.7634 & 11.9159 \\
 0.15 & 4.7674 & 7.0788 & 9.5109 & 11.6495 & 4.8933 & 7.2409 & 9.6760 & 11.8015 \\
 0.07 & 4.7278 & 7.0162 & 9.4401 & 11.5695 & 4.7455 & 7.0610 & 9.4786 & 11.6114 \\
 0.05 & 4.7176 & 7.0017 & 9.4224 & 11.5420 & 4.7021 & 6.9975 & 9.4308 & 11.5441 \\
 0.02 & 4.6957 & 6.9712 & 9.3860 & 11.4940 & 4.6422 & 6.9144 & 9.3175 & 11.4477 \\
 0.01 & 4.6870 & 6.9581 & 9.3698 & 11.4693 & 4.6150 & 6.8633 & 9.2743 & 11.3973 \\
\hline
\multicolumn{9}{c}{\mbox{}}
\end{tabular}
\caption{Values of expected delays $\mbox{ED}$ for the CUSUM and Shiryaev methods for different $\mbox{ARL}^0 = 100,500,2\,500,10\,000$ for
a shift from encrypted to compressed data}
\label{tab:ED}
\end{table*}

However, in the case of detecting encrypted data, expected delays are less relevant as a measure of performance since the data can be handled without any time aspect: the goal is to detect accurately where the encrypted data is located. A method with long expected or conditional expected delay merely means a slightly less efficient procedure.

\begin{figure*}[t]
    \centering
    \begin{minipage}[t]{9cm}
    \includegraphics [width=\linewidth]{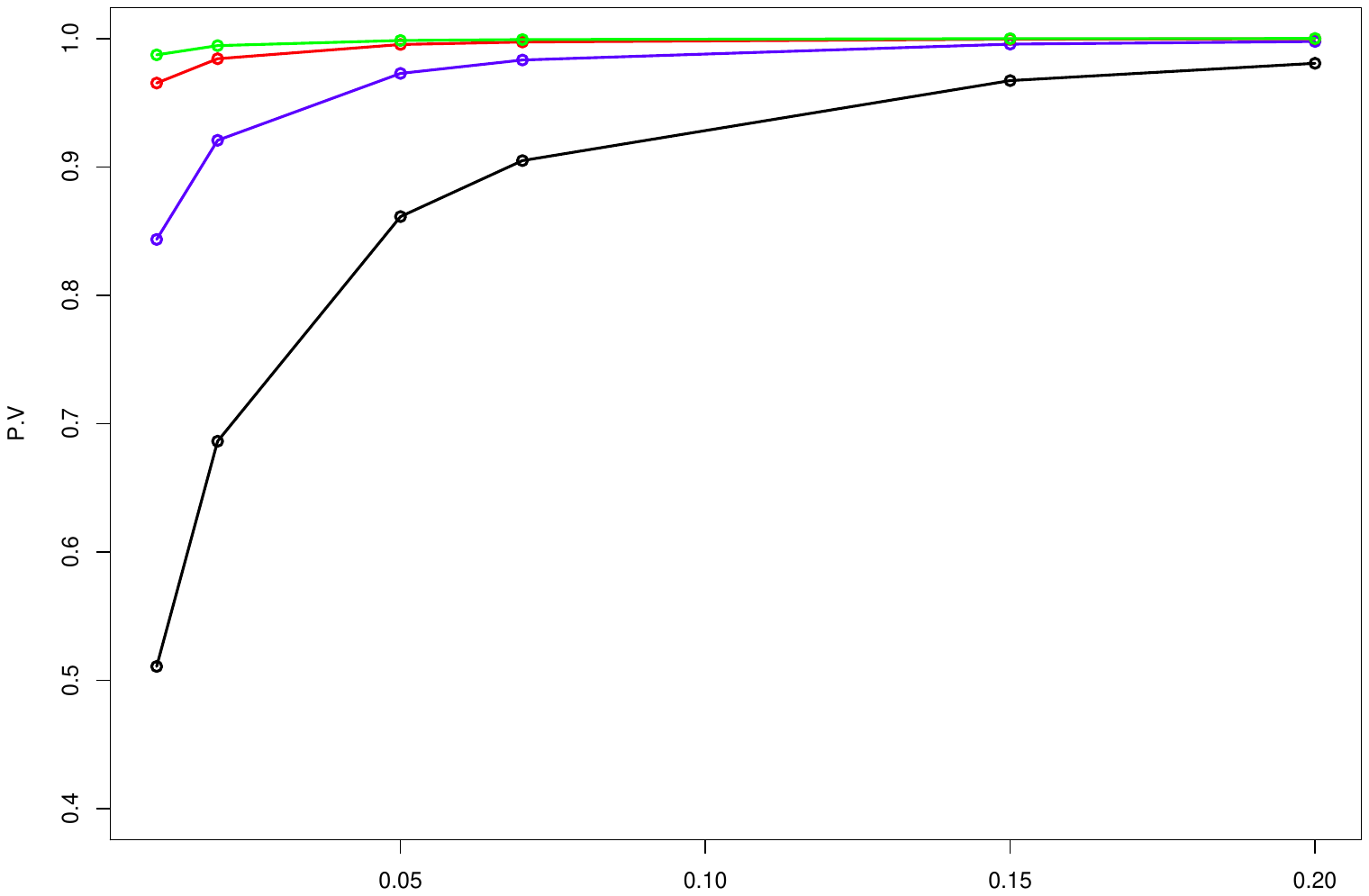}
    \put(-120,10){\footnotesize $\nu$}
    \put(-85,60){\textcolor{black}{\footnotesize $\mbox{ARL}^0 = 100$}}
    \put(-85,75){\textcolor{blue}{\footnotesize $\mbox{ARL}^0 = 500$}}
    \put(-85,90){\textcolor{red}{\footnotesize $\mbox{ARL}^0 = 2\,500$}}
    \put(-85,105){\textcolor{green}{\footnotesize $\mbox{ARL}^0 = 10\,000$}}
    \end{minipage}
    \begin{minipage}[t]{9cm}
    \includegraphics [width=\linewidth]{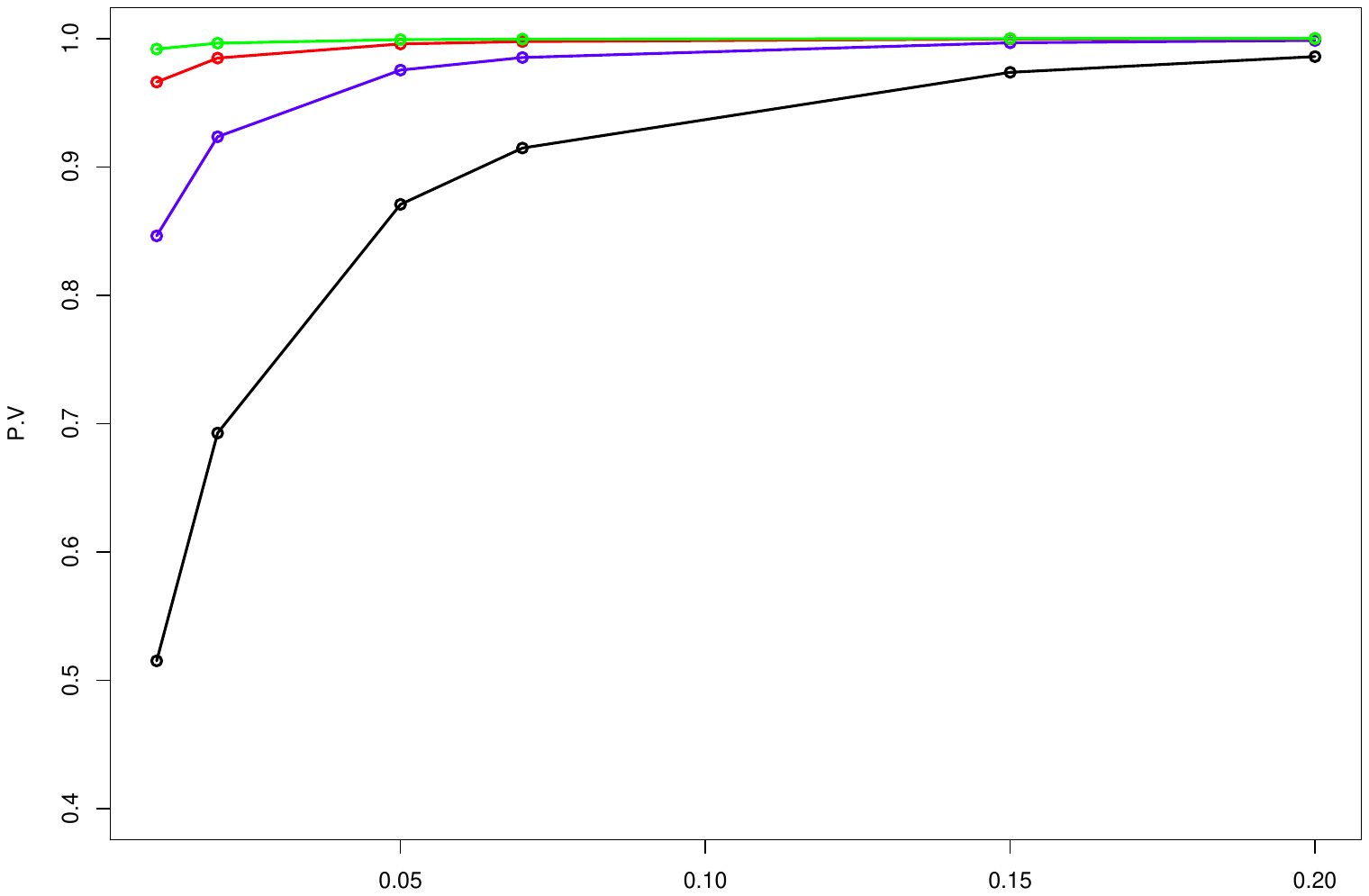}
    \put(-120,10){\footnotesize $\nu$}
    \put(-85,60){\textcolor{black}{\footnotesize $\mbox{ARL}^0 = 100$}}
    \put(-85,75){\textcolor{blue}{\footnotesize $\mbox{ARL}^0 = 500$}}
    \put(-85,90){\textcolor{red}{\footnotesize $\mbox{ARL}^0 = 2\,500$}}
    \put(-85,105){\textcolor{green}{\footnotesize $\mbox{ARL}^0 = 10\,000$}}
    \end{minipage}
    \caption{Predictive values for a shift from compressed to encrypted data for the CUSUM procedure  (left) and for the Shiryaev procedure (right)}
\label{fig:PVCUSUMDown}
\end{figure*}

A more relevant performance indicator, in this case, is for instance the predictive value $\mbox{PV}=P(\theta<\tau)$ (the probability that the method signal alarm when the change-point has actually occurred; see Table~\ref{tab:PV} and Figure~\ref{fig:PVCUSUMDown}) or the percentage of detected encrypted files that is discovered while running the process and how to improve it (see Figure \ref{tab:percentage}).

While running the process, the method will stop at some time, $\tau$, and then estimate the change point, $\theta$, by maximizing the likelihood function by using the data after the last previous alarm and the newly detected change-point. This estimated change-point, $\hat{\theta}$, can be either before or after the true change-point $\theta$. One could increase the intervals where encrypted data were discovered. This would lead to missing less encrypted data (see Table~\ref{tab:percentage}) but also brute-forcing more non-encrypted data. In Table~\ref{tab:percentage} intervals of the form $[\tau_1-i,\tau_2+i]$ are considered as estimated regions of encrypted data. Typically with large $i$, the values are very close but not exactly equal to $1$; this happens because the change points are very close (i.e.\ less than 10 clusters apart for example) and the method does not detect any change. Then no ciphertext is detected at all.

\begin{table}[htbp]
  \centering
\begin{tabular}{|r|c|c|}
 \hline
  \multirow{2}{*}{$i$} & \multicolumn{2}{c|}{Percentage} \\
  \cline{2-3}
   & CUSUM & Shiryaev \\
  \hline
  0 & 0.960254 & 0.961280 \\
  1 & 0.971053 & 0.971274 \\
  2 & 0.976242 & 0.978665 \\
  3 & 0.979266 & 0.980872 \\
  4 & 0.983681 & 0.985597 \\
  5 & 0.986248 & 0.986101 \\
  10 & 0.990101 & 0.990101 \\
  50 & 0.993371 & 0.994931 \\
  100 & 0.994326 & 0.995682 \\
 \hline
\multicolumn{3}{c}{\mbox{}}
\end{tabular}
\caption{Percentage of encrypted files that are detected when the interval of detected change points $[\tau_1,\tau_2]$ is inflated to $[\tau_1-i,\tau_2+i]$ and $i=0,1,2,3,4,5,10,50,100$.}
\label{tab:percentage}
\end{table}

Therefore the difference between the change-points and the alarms according to the method is calculated. Since the proportion of encrypted data relative to the total amount of data on the HDD is unknown, the expected proportion of error is suggested. This is to say, given two consecutive change-points, $\theta_1$ and $\theta_2$, and two corresponding stopping times, $\tau_1$ and $\tau_2$, the expected proportion of error is $E\left(\frac{|\tau_1-\theta_1|+|\tau_2-\theta_2|}{\theta_2-\theta_1}\right)$. But of course, this value has a sense only when there are no false alarms between $\tau_1$ and $\tau_2$.

If there are false alarms between $\tau_1$ and $\tau_2$, the proportion of undetected encrypted data was added to the proportion of error
%$E\left(\frac{|\tau_1-\theta_1|+|\tau_2-\theta_2|}{\theta_2-\theta_1}\right)$ were added 
to determine the proportion of the error made relative to the size of the encrypted data. Assuming that there are $n$ false alarms $\tau_1'<\ldots<\tau_n'$ in $[\tau_1,\tau_2]$, called {\em expected inaccuracy}, or $\mbox{EI}$ for short, is defined as follows.
\[ \mbox{EI}(\nu)=E_\nu\Bigg(\frac{|\tau_1-\theta_1|+|\tau_2-\theta_2|}{\theta_2-\theta_1}+\sum_{i=1}^{n/2}\frac{\tau_{2i}'-\tau_{2i-1}'}{\theta_2-\theta_1}\Bigg). \]

\begin{figure}
 \centering
 \includegraphics[width=8cm]{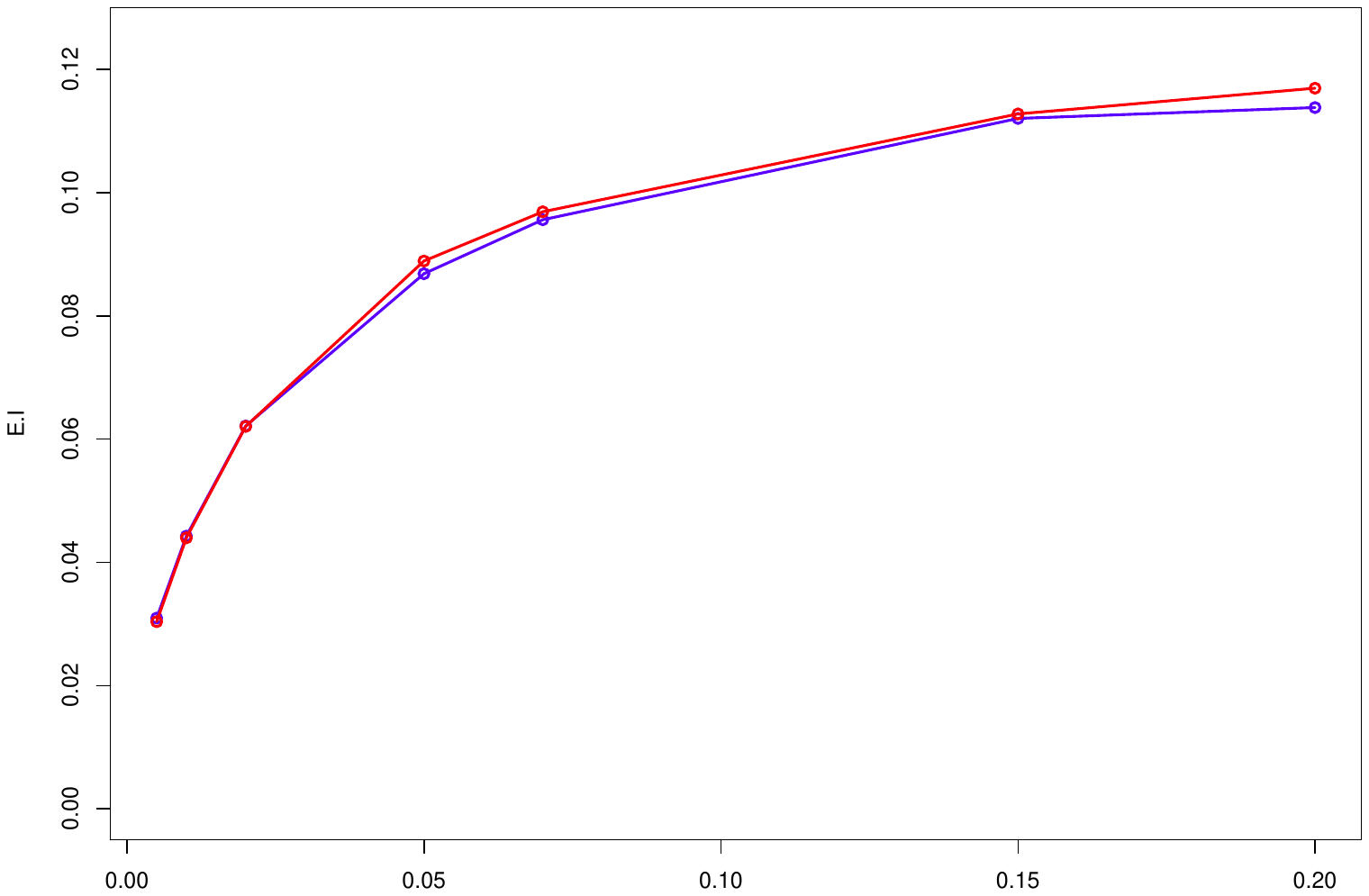}
 \put(-110,3){$\nu$}
 \caption{Expected inaccuracy $\mbox{EI}$ of the CUSUM and Shiryaev procedure. The Shiryaev method is little less accurate when
 $\nu$ increases but slightly more accurate for small $\nu$ compared to the CUSUM procedure.}
 \label{EI:plot}
\end{figure}

The $\mbox{EI}$ was measured for different values of the parameter $\nu$ in the geometrical distribution of the change-points for different methods (see Table \ref{EI} and Figure \ref{EI:plot}).

\begin{table*}[ht]
  \centering
{\footnotesize\begin{tabular}{|r|r|c|c|c|c|c|c|c|c|}\cline{2-10}
\multicolumn{1}{r|}{\mbox{}} & {$\nu$} & 0.2 & 0.15 & 0.07 & 0.05 & 0.02 & 0.01 & 0.005 & 0.001\\ \hline
\multirow{2}{*}{Percentage} & \mbox{CUSUM} & 0.11379 & 0.11202 & 0.09559 & 0.08683 & 0.06214 & 0.04426 & 0.03096 & 0.01655\\
   & Shiryaev & 0.11693 & 0.11276 & 0.09688 & 0.08890 & 0.06207 & 0.04398 & 0.03038 & 0.01603\\ \hline
\multicolumn{10}{c}{\mbox{}}
\end{tabular}}
\caption{$\mbox{EI}$ for CUSUM and Shiryaev change-point detection methods and for some values of the parameter $\nu$.}
\label{EI}
\end{table*}

%\begin{table}
%  \centering
%\begin{tabular}{|c|c|c|}
% \hline
%  \multirow{2}{*}{$\nu$} & \multicolumn{2}{c|}{Percentage} \\
%  \cline{2-3}
%   & CUSUM & Shiryaev \\
%  \hline
%  0.20 & 0.113791 & 0.116934 \\
%  0.15 & 0.112024 & 0.112757 \\
%  0.07 & 0.095589 & 0.096884 \\
%  0.05 & 0.086831 & 0.088897 \\
%  0.02 & 0.062135 & 0.062066 \\
%  0.01 & 0.044261 & 0.043975 \\
%  0.005 & 0.030960 & 0.030376 \\
%  0.001 & 0.016548 & 0.016028 \\
% \hline
%\end{tabular}
%\caption{$EI$ for different parameters $\nu$ and different methods}
%\label{EI}
%\end{table}

\paragraph*{Complete procedure}
The complete procedure returns a segmentation separating suspected encrypted data and most likely non-encrypted data of an HDD, information provided in order to carry out the subsequent brute force cryptanalysis efficiently. This procedure runs a likelihood ratio based change-point detection method and as soon as it detects a change, maximizes a likelihood function to find the most likely estimator of the change-point to determine where the real change is most likely located. It will then start over from the location of this estimated change-point with the same method for on-line change-point detection except that the likelihood ratio is reversed modifying the alarm function to fit with the opposite change-point situation, and so on.

\section*{Results}
\paragraph*{Thresholds and experimental values for the method}
The first step in establishing properties of change-point detection methods is to determine the thresholds rendering values of average run-length, $\mbox{ARL}^0$. One purpose of this is simply to link the threshold values to a property which relates to the probability of false alarm. This is a result in itself, but it is also necessary for calibrating the methods so they are comparable in terms of other performance measures related to a motivated alarm, such as expected delay, predictive value and expected inaccuracy. Here the values $\mbox{ARL}^0=100,500,2\,500$ and  $10\,000$, are considered for both the CUSUM and Shiryaev methods, for a shift from encrypted to non-encrypted and vice versa. The change-points are commonly geometrically distributed with parameter $\nu$. Here the average time before a change-point is expected to be rather high (several hundreds or thousands maybe) as the method deal with 64-byte clusters in an HDD of surely several hundreds of Giga or Tera Bytes. Thus, since $E(\tau)=1/\nu$, the focus is on very small values of $\nu$ for the methods to be sensibly trigger happy.

\begin{table*}[t]
  \centering
\begin{tabular}{|r|l|l|l|l|}
 \hline
 \multirow{3}{*}{$\mbox{ARL}^0$} & \multicolumn{4}{c|}{Thresholds} \\
 \cline{2-5}
  & \multicolumn{2}{c|}{CUSUM} & \multicolumn{2}{c|}{Shiryaev} \\
 \cline{2-5}
  & $\mbox{NE}\rightarrow\mbox{E}$ & $\mbox{E}\rightarrow\mbox{NE}$ & $\mbox{NE}\rightarrow\mbox{E}$ & $\mbox{E}\rightarrow\mbox{NE}$ \\
 \hline
 100  & 1.2260 & 4.5801 & 64.0313 & 44.1271   \\
 500 & 2.6529 & 6.1250 & 323.0625 & 221.8125  \\
 2\,500 & 4.2188 & 7.7120 & 1618.219 & 735.4088   \\
 10\,000 & 5.5296 & 9.0990 & 6475.0547 & 4441.8413 \\
\hline
\multicolumn{5}{c}{\mbox{}}
\end{tabular}
\caption{Values of the thresholds for the CUSUM and Shiryaev methods for $\mbox{ARL}^0 = 100,500,2\,500,10\,000$ specified for detecting a shift from non-encrypted to encrypted data (indicated by $\mbox{NE}\rightarrow\mbox{E}$ for short) and for shift from encrypted to non-encrypted data (indicated by $\mbox{E}\rightarrow\mbox{NE}$) respectively.}
\label{tab:threshold}
\end{table*}

\begin{figure*}[t]
\centering
\includegraphics[width=\linewidth]{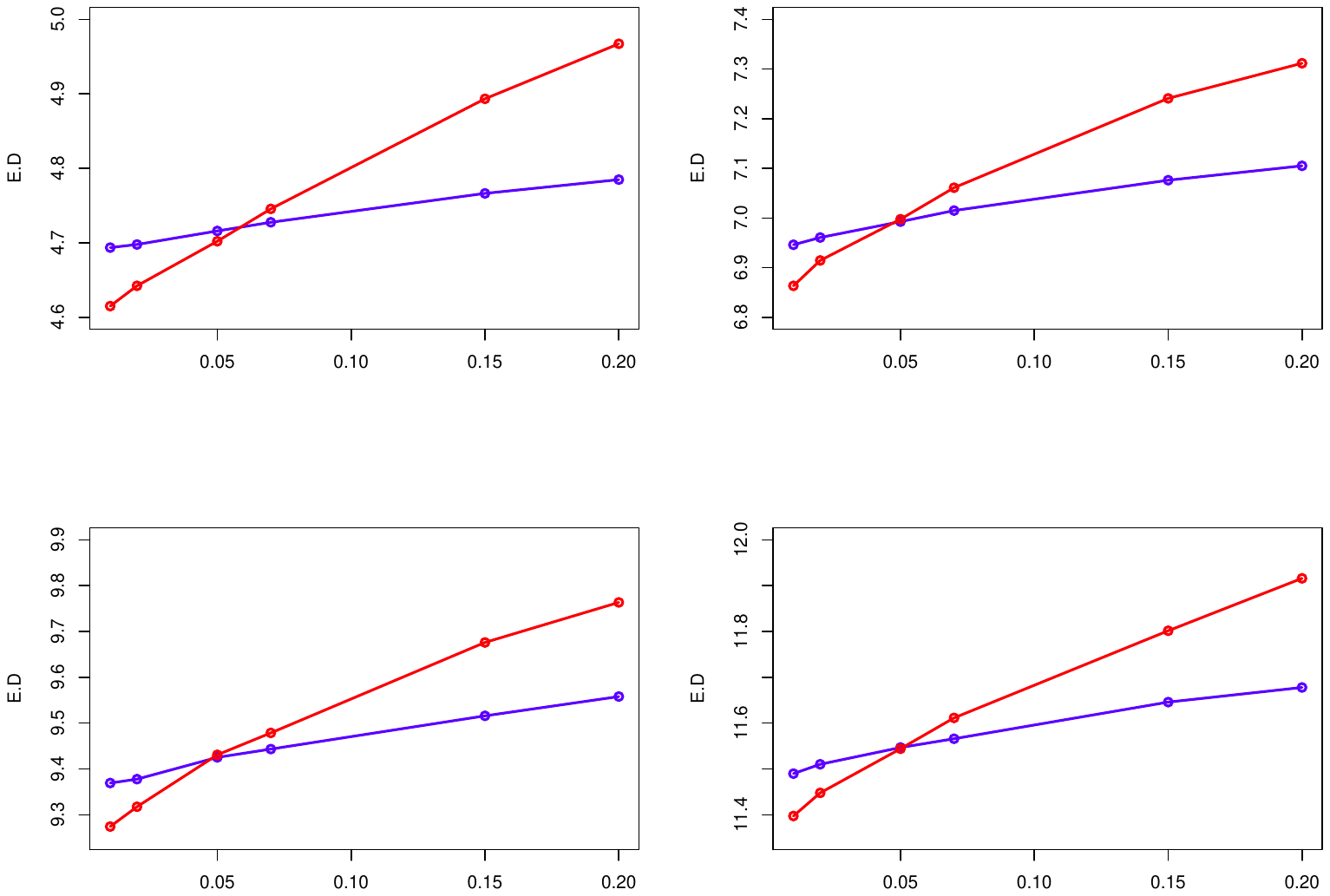}
\put(-425,390){\footnotesize $\mbox{ARL}^0 = 100$}
\put(-160,390){\footnotesize $\mbox{ARL}^0 = 500$}
\put(-425,190){\footnotesize $\mbox{ARL}^0 = 2\,500$}
\put(-160,190){\footnotesize $\mbox{ARL}^0 = 10\,000$}
\put(-400,25){\footnotesize $\nu$}
\put(-400,225){\footnotesize $\nu$}
\put(-135,25){\footnotesize $\nu$}
\put(-135,225){\footnotesize $\nu$}
\caption{Expected delays, $\mbox{ED}$, for a shift from encrypted to compressed data for the CUSUM procedure (blue) and Shiryaev procedure (red)}
\label{fig:EDup}
\end{figure*}

Commonly values of $\mbox{ARL}^0$ are 100 or 500 in order to make other properties relevant for comparisons. In the case of this application, however, as large values of $\mbox{ARL}^0$ as $2\,500$ and $10\,000$ are studied because the first change-point might not occur until far into the HDD. Adjusting the threshold by simulating data can take a very long time if $\mbox{ARL}^0$ is large ($2\,500$ or $10\,000$, especially for the Shiryaev method). In this case, it can take several hours or even up to days to compute the threshold. Therefore it would be interesting to have a way of predicting the threshold by extrapolation i.e.\  having an explicit relation between $\mbox{ARL}^0$ and the threshold $C$. Intuitively, if $\mbox{ARL}^0$ is larger, more data will be taken into account implying a threshold proportionally larger. Indeed when $\mbox{ARL}^0$ increases more data is used in the procedure and the threshold is therefore proportionally increased from how much more data that was treated in the procedure. In the CUSUM case, since the alarm function is defined by means of the log likelihood ratio, the relationship between threshold $C$ and $\mbox{ARL}^0$ is logarithmic:
\begin{itemize}
\item for a shift from encrypted data to non-encrypted data:
\[ C = 0.997767 \cdot \ln\left(0.912316\cdot \mbox{ARL}^0 + 7.294950\right) \]
\item for a shift from non-encrypted data to encrypted data:
\[ C = 0.965524 \cdot \ln\left(0.030655\cdot \mbox{ARL}^0 + 0.494603\right) \]
\end{itemize}

For Shiryaev, the threshold $C$ is a linear function of $\mbox{ARL}^0$:
\begin{itemize}
\item for a shift from encrypted data to non-encrypted data:
\[ C = 0.444214 \cdot \mbox{ARL}^0 + 0.294281 \]
\item for a shift from non-encrypted data to encrypted data:
\[ C = 0.647578 \cdot \mbox{ARL}^0 - 0.726563 \]
\end{itemize}

\begin{table*}[t]
  \centering
\begin{tabular}{|c|c|c|c|c|c|c|c|c|c|}
 \hline
 \multirow{3}{*}{$\nu$} & \multicolumn{8}{c|}{Methods} \\
 \cline{2-9}
  & \multicolumn{4}{c|}{CUSUM} & \multicolumn{4}{c|}{Shiryaev} \\
 \cline{2-9}
  & 100 & 500 & 2\,500 & 10\,000 & 100 & 500 & 2\,500 & 10\,000 \\
 \hline
 0.20 & 0.8950 & 0.9862 & 0.9984 & 0.9997 & 0.9859 & 0.9984 & 0.9998 & 1.0000 \\
 0.15 & 0.8727 & 0.9805 & 0.9974 & 0.9995 & 0.9736 & 0.9967 & 0.9996 & 0.9999 \\
 0.07 & 0.8031 & 0.9604 & 0.9933 & 0.9986 & 0.9147 & 0.9852 & 0.9976 & 0.9995 \\
 0.05 & 0.7634 & 0.9482 & 0.9907 & 0.9978 & 0.8708 & 0.9754 & 0.9957 & 0.9991 \\
 0.02 & 0.6171 & 0.8942 & 0.9784 & 0.9944 & 0.6927 & 0.9235 & 0.9847 & 0.9964 \\
 0.01 & 0.4712 & 0.8207 & 0.9590 & 0.9890 & 0.5153 & 0.8463 & 0.9661 & 0.9918 \\
\hline
\multicolumn{9}{c}{\mbox{}}
\end{tabular}
\caption{Predictive value $\mbox{PV}(\nu)=P_\nu(\theta<\tau)$, i.e.\ the probability that a shift has occurred when an alarm is signalled, for the CUSUM and the Shiryaev methods, for different values of $\mbox{ARL}^0$ and with different values of the parameter $\nu$ in the geometric distribution of the change-points.}
\label{tab:PV}
\end{table*}

\section*{Conclusions}
Using the change-point theory, methods to detect encrypted data on HDDs were successfully derived and evaluated. These methods exploit the fact that encrypted data is uniformly distributed as opposed to other types of files. The methods were designed to detect the difference between encrypted and non-encrypted data where the kind of non-encrypted data that was most similar to the encrypted data was compressed data. As the proposed methods detect even this small a difference in the data, any bigger deviance will be even easier detected.  % Therefore this process is likely to detect encrypted data among any type of data.
%In contrast, if the device contains only 0s (is empty in some sense) on an interval it will maximise the $u$-value which will attain 448.0 and this part will not be regarded as encrypted data because it is a high value.

Quick and accurate detection of a change is commonly the desired property of change-point detection methods. In many applications, time aspects of the methods are a matter of interest, e.g.\ expected delay in detection of a shift or probability of detecting a shift within a specified time interval. 
Here, however, this time aspect is not of primary interest since the data remain the same during the whole process. Here the probability of correctly detecting encrypted data is more relevant. The investigation shows that the proposed methods detect more than 96\% of the encrypted data and, by extending the intervals, the methods detect more than 99\% of the encrypted data. By assuming that the change-points are not too close -- which is a plausible assumption since it is unlikely that files are so small if the device is not too fragmented -- then the method, by adding a little margin to the intervals, quickly detects 100\% of the encrypted data.

The change-point itself is a random variable. However, for some of the performance measures, the results depend on whether the change is more likely to occur early or late. In the application of detecting encrypted data the distance between the shift from encrypted to non-encrypted data and vice versa is typically longer than 20 clusters which corresponds to parameter value $\nu<0.05$ in the geometrical distribution of the change-point. Therefore these values are more interesting in the case when the hard drive is not very strongly fragmented. Then the Shiryaev method turns out to be slightly better compared to the CUSUM method in respect of expected delay. The Shiryaev method also detects more encrypted data than the CUSUM method and has a slightly higher predictive value $\mbox{PV}$.

All in all, this implies that both methods designed with the suggested modelling, perform very well with a slight preference to the Shiryaev method for detecting encrypted data on an HDD.

\section*{Acknowledgements}
The authors wish to express their gratitude to Mattias Weckst\'{e}n at Halmstad University for good ideas and previous readings of the manuscript University and to Linus Nissi (formerly Linus Barkman) at the Police Department of Southern Sweden for earlier work in the area.

%\externalbibliography{yes}
%\bibliographystyle{IEEEtran}
%\bibliography{IEEEabrv,ref.bib}

\end{document}